%% file: main.tex
\documentclass[acmtog]{acmart}

\acmSubmissionID{396}

\usepackage{amsmath}
\usepackage{algorithmic}
\usepackage[ruled,vlined]{algorithm2e}
\usepackage{bbm}
\usepackage{subfigure}

\input{defs}
\setcopyright{acmcopyright}
\acmJournal{TOG}
\acmYear{2020}
\acmVolume{39}
\acmNumber{6}
\acmArticle{189}
\acmMonth{12} 
\acmDOI{10.1145/3414685.3417868}

\begin{document}



\title{Learning to Manipulate Amorphous Materials}


\author{Yunbo Zhang}
\affiliation{\institution{Georgia Institute of Technology, USA}}
\author{Wenhao Yu}
\affiliation{\institution{Georgia Institute of Technology, USA}}
\author{C. Karen Liu}
\affiliation{\institution{Stanford University, USA}}
\author{Charles C. Kemp}

\affiliation{\institution{Georgia Institute of Technology, USA}}
\author{Greg Turk}
\affiliation{\institution{Georgia Institute of Technology, USA}}


\begin{abstract}


We present a method of training character manipulation of amorphous materials such as those often used in cooking.  Common examples of amorphous materials include granular materials (salt, uncooked rice), fluids (honey), and visco-plastic materials (sticky rice, softened butter).  A typical task is to spread a given material out across a flat surface using a tool such as a scraper or knife.  We use reinforcement learning to train our controllers to manipulate materials in various ways.   The training is performed in a physics simulator that uses position-based dynamics of particles to simulate the materials to be manipulated.  The neural network control policy is given observations of the material (e.g. a low-resolution density map), and the policy outputs actions such as rotating and translating the knife.  We demonstrate policies that have been successfully trained to carry out the following tasks: spreading, gathering, and flipping.  We produce a final animation by using inverse kinematics to guide a character’s arm and hand to match the motion of the manipulation tool such as a knife or a frying pan.

\end{abstract}



\begin{teaserfigure}
    \centering
    \includegraphics[height=2.4cm]{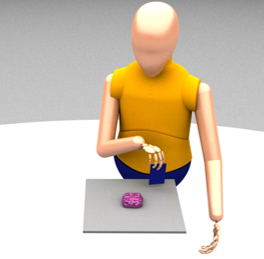}
    \hfill
    \includegraphics[height=2.4cm]{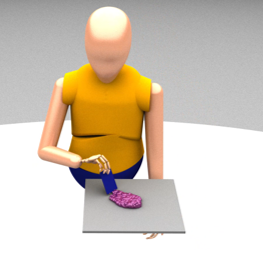}
    \hfill
    \includegraphics[height=2.4cm]{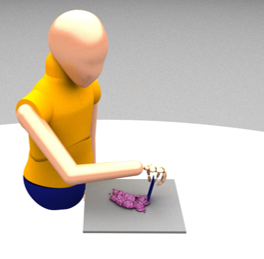}
    \hfill
    \includegraphics[height=2.4cm]{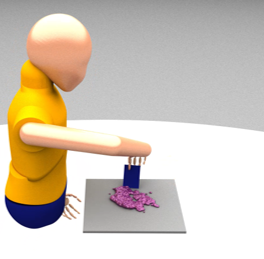}
    \hfill
    \includegraphics[height=2.4cm]{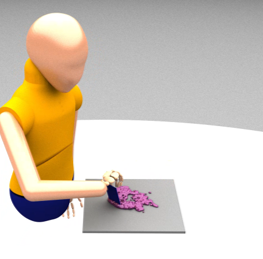}
    \hfill
    \includegraphics[height=2.4cm]{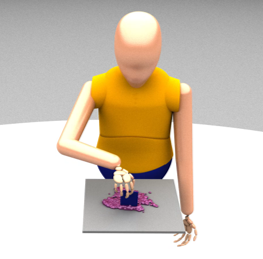}
    \hfill
    \includegraphics[height=2.4cm]{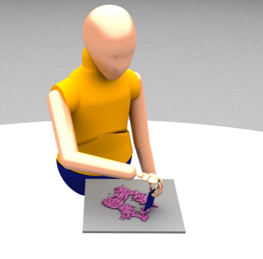}
    \caption{A character is controlling a scraper to spread visco-plastic material.}
  \label{fig:teaser}
\end{teaserfigure}

\maketitle

\input{introduction.tex}

\input{related_work.tex}
\input{methods.tex}

\input{results.tex}
\input{ablation_study.tex}

\input{generalization.tex}

\input{discussion.tex}
\input{conclusion.tex}

\begin{acks}
We thank Alexander Clegg, Henry Clever, Zackory Erickson and Yifeng Jiang for helpful discussions and technical assistance. This work is supported by NSF award IIS-1514258.
\end{acks}



\bibliographystyle{ACM-Reference-Format}
\bibliography{references}



\end{document}

%% file: defs.tex
\newcommand{\cmt}[1]{}

\long\def\ignorethis#1{}

\newcommand{\etal}{{\em{et~al.}\ }}

\newcommand{\argmax}{\operatornamewithlimits{argmax}}

\newcommand{\pctab}{\hspace{0.2in}}

\newcommand\high{0.9in}



%% file: introduction.tex
\section{Introduction}

People can manipulate materials that have a wide variety of physical properties, and we would like our virtual characters to be similarly skilled at manipulation.  One broad class of materials that has received little attention for character animation are materials that have no fixed shape, so-called \emph{amorphous} materials.  Amorphous materials include granular materials, fluids, and visco-plastics.  Our goal is to develop character control policies for manipulating such amorphous materials.  In day-to-day living, the foods that we cook in our kitchen provide numerous examples of such materials.  Examples of manipulating amorphous materials in the kitchen include spreading butter on toast, gathering fried rice into a pile to be served, and flipping pancakes or burgers on a skillet.

Rather than creating hand-written controllers for our characters, we draw upon the strengths of reinforcement learning algorithms to develop robust character control policies. Researchers have successfully applied reinforcement learning to create policies that control characters to perform a variety of motor skills such as parkour \cite{peng2018deepmimic}, basketball dribbling \cite{liu2018learning}, ice skating \cite{yu2019figure} and dressing \cite{clegg2018learning}. 

We separate the character control task into two parts. First, we use reinforcement learning to train a material manipulation policy (e.g. spreading rice) that describes the motion of a tool such as a knife or a frying pan. The character’s body is not considered during the training of the tool motion. Second, when producing a final animation, we use inverse kinematics to guide the arm and hand of a character so that it matches the motion of the tool as the material is being manipulated.


Numerous character animation controllers have been created for picking up, moving and placing rigid objects. When interacting with such a discrete rigid object, the object’s position and orientation require just six numbers.  In contrast, creating observations of amorphous materials for a character requires a more verbose object description.  Consider, for example, rice that is on the heated surface that is being cooked by a hibachi chef.  The rice might be scattered across the heat surface or may be heaped in one or more piles.  When manipulating such materials, our animated characters need to be able to recognize and act in different ways based on the rice’s configuration.  We have chosen to describe a given material using either a material density map or a depth image, which gives the character policy a rich description of the configuration of the material.  Furthermore, we find that the coordinate frame of these image maps has a large effect on the speed at which a controller will learn a task.  In particular, we find that for the tasks we investigated, tool-centric observations are superior to observations in world-space.

Another way in which characters acting upon amorphous materials differs from other animation tasks is the quantitative evaluation of a given action.  When using reinforcement learning for common tasks such as locomotion, the general nature of a reward function is well known.  A typical reward function might be a term for distance travelled, a balance term, and an energy term.  Similarly, reward terms for reaching out a robotic arm to a given object are commonplace.  However, we have found no such ready-made rewards in the research literature for tasks such as scattering or gathering granular materials.  Much of our effort has been devoted to discovering effective ways in which to reward a given character policy for such tasks.

A challenge to using reinforcement learning for amorphous material manipulation is that simulating amorphous materials comes at a high computational cost.  We address this problem by using the FleX simulator, which is capable of simulating a wide range of materials using a position-based particle model~\cite{macklin2014unified}.  When simulating the manipulation of materials, we use FleX to perform several rollouts simultaneously in a single simulation environment on the GPU, similar to the work of Chebotar \etal on robot training in simulation~\shortcite{chebotar2019closing}. 

%% file: related_work.tex
\section{Related Work}


Controlling simulated characters and robots to exhibit complex and versatile manipulation skills is a long-standing challenge in computer animation, robotics, and machine learning. Researchers have investigated a large variety of manipulation problems that span different types of objects and end effectors such as object grasping \cite{liu2009dextrous, andrews2012policies, DmitryQT, zhao2013robust}, in-hand manipulation \cite{bai2014dexterous, mordatch2012contact, akkaya2019solving, ye2012synthesis}, cloth folding and manipulation \cite{bai2016dexterous, miller2012geometric, wu2019learning}, and scooping \cite{schenck2017learning, park2019active}. Many of these works leverage trajectory optimization algorithms to synthesize the manipulation motion. For example, Bai \etal developed an algorithm to find a trajectory where a piece of cloth or garment was manipulated into desired states through contact forces \shortcite{bai2016dexterous}. However, these methods are usually specific to a particular motion and require frequent re-planning of the motion in order to combat uncertainties and perturbations in the environment.

Deep reinforcement learning (DRL) provides a general framework for characters to automatically find a control policy from rewards that can effectively handle uncertainties and perturbations in the system. Recent advancements in DRL algorithms have enabled researchers to create control policies for a variety of character motor skills \cite{peng2018deepmimic, liu2018learning, yu2019figure, lee2019scalable}, including ones for manipulation problems such as solving rubik's cube \cite{akkaya2019solving} and opening doors \cite{rajeswaran2017learning}. However, due to the high sample complexity of DRL algorithms and the high computation time for non-rigid objects like fluids and cloths, it is most common to apply DRL algorithms to manipulation problems with rigid objects, for which stable and efficient simulation tools are available \cite{todorov2012mujoco, coumans2016pybullet, lee2018dart}. For the work that do apply DRL algorithms to manipulating non-rigid objects, some form of prior knowledge about the task or the object is usually utilized to make the problem more tractable \cite{elliott2018robotic, clegg2018learning, ma2018fluid, wu2019learning, wilson2019learning}. For example, Clegg \etal applied DRL to teach a simulated humanoid character to dress different garments \shortcite{clegg2018learning}. To make the dressing problem manageable, they decomposed the dressing motion into a set of more manageable sub-tasks, which requires manual effort for new dressing problems. More recently, Wu \etal proposed a DRL-based approach for manipulating cloth and ropes from visual input \shortcite{wu2019learning}. They assumed the robot to alternate between picking and placing the object during the manipulation and developed a learning algorithm that is suitable for this action space. 

Ma \etal proposed a learning system for using fluid to control rigid objects \shortcite{ma2018fluid}. They learned an auto-encoder for extracting compact features from the input image, which is used as the observation space in training the control policy. Our work can be viewed as solving an opposite problem: we actuate a rigid object such as a pan or a spatula to manipulate the state of amorphous materials. We do not use an auto-encoder stage in our network architecture, but instead use a two-stage convolutional neural network to process the images that represent the state of the materials.


Another way to deal with the high computation complexity of non-rigid objects for efficient learning is to use a differentiable representation of the objects' dynamics. Researchers have proposed to represent the dynamics model using deep neural networks \cite{schenck2018spnets, schenck2017learning, li2018learning}. For instance, Schenck \etal formulated the position-based dynamics algorithm as neural network operations that can be embedded inside a bigger neural network \shortcite{schenck2018spnets}. Li \etal developed a graph network-based approach for representing the dynamics of particle-based objects \shortcite{li2018learning}. They show their method on manipulating fluid and deformable objects such as making a rice ball.  The success of these methods usually relies on being able to collect training data that are relevant to the task of interest, limiting them to relatively simple forms of tool-object interactions. Alternatively, researchers have also investigated formulating and implementing physics of these objects in a differentiable way \cite{hu2019chainqueen, hu2019difftaichi}. These methods have great potential in achieving effective learning for complex materials. However, it can be non-trivial and requires human expertise to create such a model for new problems that involve novel materials and interactions. 


%% file: methods.tex
\section{Methods}

In this section, we present our pipeline for training policies that control tools to manipulate amorphous materials. We will first give an overview of the reinforcement learning control problem formulation. We will then introduce our representation for the material states and discuss our controller design. We next discuss policy training details and our simulation environment. Lastly, we describe how we generate a character animation from a trained policy.

\subsection{Policy Training Formulation}

We use reinforcement learning to train our control policies for manipulation. We use a neural network to represent the policy, and this network's input is the observed state of the material and the manipulation tool, and the output is the tool motion (the action). To train the policy, we run many different trajectories, and after each action a reward is given that is used to tune the weights of the neural network.

More formally, we formulate our control problem as a Markov Decision Process (MDP) defined by the tuple $\left( \mathcal{S},\mathcal{A},\mathcal{T},r,\rho_{0},\gamma \right)$. $\mathcal{S}$ is the state space of the environment, $\mathcal{A}$ is the action space, $\mathcal{T}(s_t,a_t) = s_{t+1}$ is the transition function (the simulation of the environment) that generate a new state $s_{t+1}$ given a state action pair $(s_t,a_t)$, $r(s_t,a_t)$ is the reward function evaluating the quality of a state action pair, $\rho_{0}$ is the initial state distribution, and $\gamma \in \left[ 0,1\right]$ is the discount factor. The goal is to find an optimal control policy $\pi_{\theta}$, that by solving for $\theta$, maximizing the following sum of expected rewards over a distribution of all trajectories $\tau = (s_0,\pi(s_0),\dots,\pi(s_{T-1}),s_T)$.
\begin{equation}
    R(\theta) = \mathbb{E}_{\tau}\left[\sum_{t=0}^{T}\gamma^{t}r(s_t, a_t)\right]
\end{equation}


\subsection{Material State Representation}

The state space $\mathcal{S}$ of our problem contains information of the controlled tool as well as the state of the amorphous material. Training the control policy with full state information of the material as input, e.g. positions and velocities of all the particles in a granular material, is not practical due to the high dimensionality of the material state. It also limits the policy to a fixed material setting, e.g. the same number of particles. Using the image of the scene as input can allow the policy to generalize to different amounts of materials, yet it does not generalize to similar materials but with different appearances. In this work, we represent the material state for input to the policy using a 2D grid of values that summarize the density and/or height of the simulated materials at each grid point. This allows the policy to quickly adapt to novel materials with minimal change. 


\subsubsection{Density Map}

In our work, we represent the simulated material as particles. We denote the position of each particle as: $\mathbf{x}_{mat}^{p}$, where $p \in \{0, 1, \cdots, P \}$ is the index of the particle among the total $P$ particles. To estimate the density of the material, we take an Eulerian view by discretizing the 2D ground plane into an $N \times N$ grid with cell width $h$ and store density information on the grid as $D \in \mathbb{R}^{N \times N}$, where each value $D(i, j)$ on the grid measures the density of material at the position of grid cell $(i,j)$. We denote the position of grid cell $(i, j)$ as $\mathbf{x}_{grid}^{i, j}$. We further write the vector from a grid cell center $(i, j)$ to a particle $p$ as $\mathbf{v}^{p,i,j} = \mathbf{x}_{mat}^{p} - \mathbf{x}_{grid}^{i, j}$. We compute the density at a certain point $D(i,j)$ as:

\begin{align}
D(i,j) = &\sum_{k \in K(i,j)} w(k, i, j),\\
w(k, i, j) = &\sqrt{(2.5h-|\mathbf{v}^{k,i,j}[0]|)^2+(2.5h-|\mathbf{v}^{k,i,j}[1]|)^2},
\end{align} where $K(i,j)$ is the set of indices for the particles that lie within the grid cell $(i,j)$ and its two-ring neighbors, i.e. $|\mathbf{v}^{k,i,j}[d]| < 2.5h$ for $ d \in \{0,1\}$. 

\subsubsection{Height Map}
Similar to the the density map, we approximate the average height of the material using a 2D height map $H \in \mathbb{R}^{N \times N}$. Using the same weight function $w$ defined above, the height value at each grid cell $(i, j)$ is computed as $H(i, j) = \frac{\sum_{p \in K(i,j)} y^p \cdot w(p, i, j)}{D(i, j)}$, where $y^p$ is the height of the $p_{\text{th}}$ particle.

\subsubsection{Relative State Representation}
\begin{figure}
    \centering
    \includegraphics[width=3.4in]{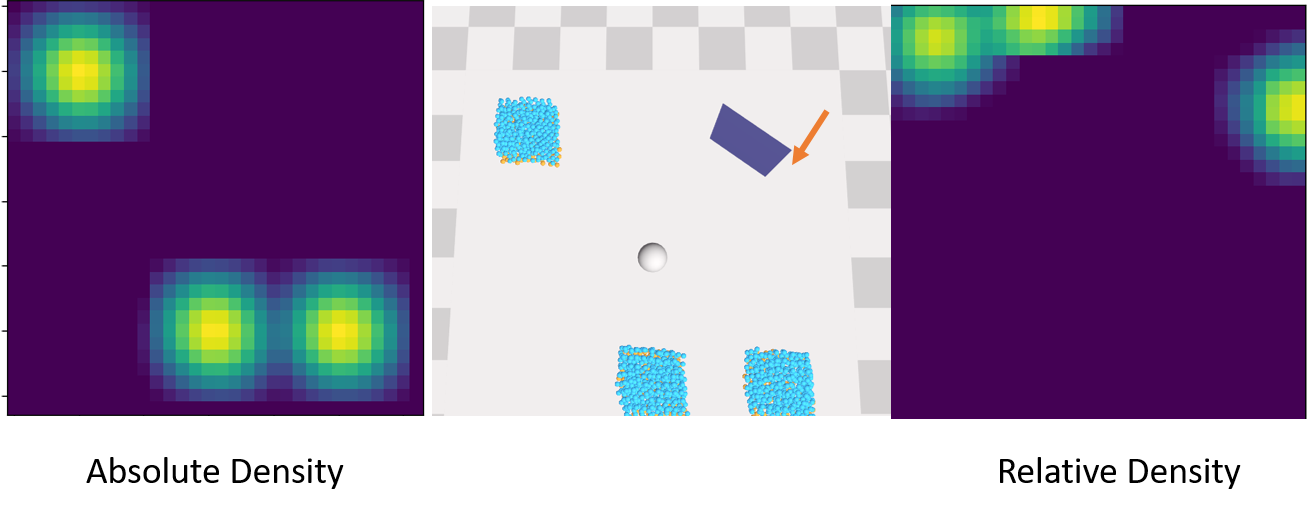}
    \vspace{-8mm}
    \caption{The middle image shows the position of tool and the facing direction indicated as a orange arrow. The left image is the corresponding density map of the material in global coordinate frame and the right image if the density map in tool's coordinate frame.}
    \vspace{-8mm}
    \label{fig:relative_obs}
\end{figure}

In many manipulation tasks it is more important to reason about the relative pose between objects of interest rather than their absolute locations in the global coordinate. For example, when assembling two lego pieces, it is crucial that they are aligned relative to each other, and it is less important what orientation or position they are in. Following this intuition, we propose to represent the state of the material in the local coordinate frame of the tool. We first extract the pose of the tool as a homogeneous transformation matrix $A$ from the global coordinate frame. We then transform the positions of all the particles in the material into the local space of the tool: $\mathbf{\tilde{x}}_{mat}^p = A^{-1} \mathbf{x}_{mat}^p$. Then we use the local particle positions to compute the density map $D$ and height map $H$ as shown in Figure \ref{fig:relative_obs}.

\subsection{Controller Design}
\subsubsection{Observation Space}

We construct our observation as $O=[O_{mat},O_{tool}]$, where $O_{mat} \in \mathbb{R}^{N \times N \times M}$ includes the material information as well as certain task-related information such as goal location. The number of input channels $M$ depends on the task and is discussed in more detail in Section \ref{section:result}. $O_{tool}$ contains the position and orientation information of the tool defined as $O_{tool} = [\mathbf{x}_{tool},\cos(\mathbf{r}_{tool}),\sin(\mathbf{r}_{tool}),\dot{\mathbf{x}}_{tool},\cos(\dot{\mathbf{r}}_{tool}),\sin(\dot{\mathbf{r}}_{tool})]$, where $\mathbf{x}_{tool}$ is the position of the tool and $\mathbf{r}_{tool}$ is the Euler angles representing the orientation of the tool. The dimensions of $\mathbf{x}_{tool}$ and $\mathbf{r}_{tool}$ depends on the tool and action space used for the task. For the angular velocity information, we use its sine and cosine value. There is a unique pair of values for the sine and cosine of the angular velocity because the angular velocity of the tool always stays in the range of $\left(-\pi,\pi\right]$ in the simulations.

\subsubsection{Action Space}

Since the tool only affects material in its vicinity, we design the action space to be the target positions and rotation angles of the tool in the tool's coordinate frame instead of the global coordinate frame. This action space design, coupled with the relative observation described above produce higher performing policies for our manipulation tasks.



%
\subsubsection{Controller Structure}

\begin{figure}
    \centering
    \includegraphics[width=3.4in]{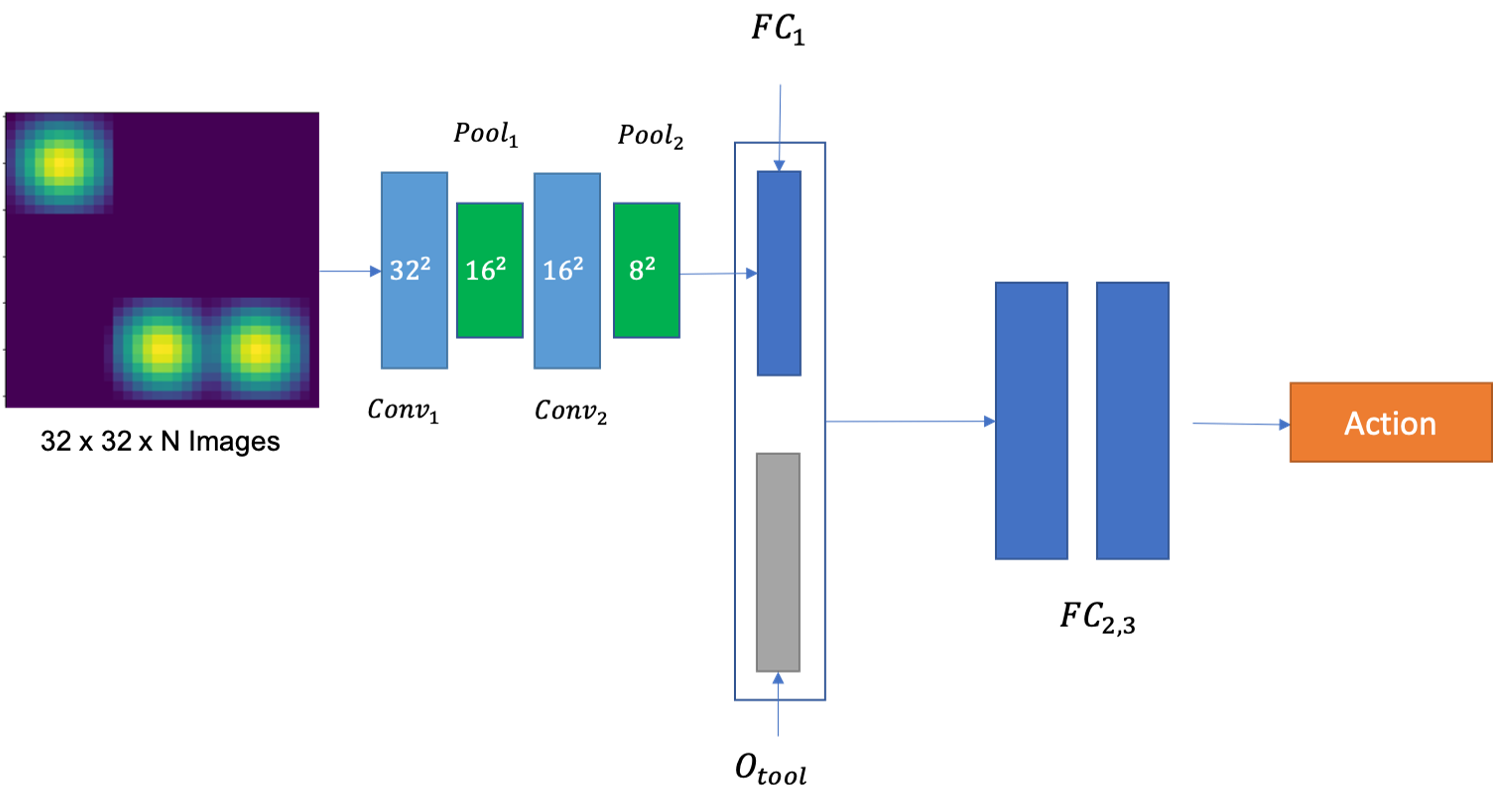}
    \vspace{-8mm}
    \caption{The convolution layers of the network are specified as: $conv_1(5,2,1)$, $conv_1(3,2,1)$ with the order of (kernel size, number of filters, stride). The pooling layers are specified identically as $pool_{1,2}(2)$, which down-samples the input with stride of $2$. The fully connected layers are specified with number of hidden units as $FC_1(32)$, $FC_2(64)$, and $FC_3(64)$.}
    \vspace{-6.5mm}
    \label{fig:network}
\end{figure}

Our observation space is a combination of 2D images and accurate tool state, thus it is natural in the design of the policy architecture to extract features of the two parts separately using different neural network blocks. For the image-based observation part, convolution layers are well suited to extract lower dimensional features. For the tool observation part, fully connected layers are used to extract the information. Hence, our neural network controller is designed as shown in Figure \ref{fig:network}. To extract the low dimensional feature from the images, we pass several $32 \times 32$ images through two convolutions layers with max pooling followed by a single fully connected layer with $32$ hidden units. The resulting feature vector is then concatenated with $O_{tool}$ and passed into two fully connected layers each with 64 hidden units. To encourage identification of positional information in the input images, our first convolution layer $conv_1$ uses the CoordConv layer design proposed in \cite{liu2018intriguing}. We used the ReLU functions as our activation function for all output of convolution layers and fully connected layers.

\subsection{Policy Training}
In this work, we use the proximal policy optimization (PPO)~\cite{schulman2017proximal} to optimize our control policy. As an on-policy algorithm, by sampling trajectories using the current policy, PPO aims to find an improved policy by optimizing a surrogate loss function as well as trying to keep the KL-divergence between the current and the previous policies within the trust region. The combined optimization loss is defined in the following form:

\begin{align}
    L_{PPO}(\theta) &=L_{Surr}+L_{KL}\nonumber\\
    &=-\mathbb{E}_{t}\left[min(r(\theta)A_t,clip(r(\theta),1-\epsilon,1+\epsilon)\right]\\
    &\ \ \ - \beta\mathbb{E}_{t}\left[KL\left[\pi_{\theta}(\cdot|s_t)|\pi_{\theta_{old}}(\cdot|s_t)\right]\right],\nonumber
\end{align}
where $r_{\theta}=\frac{\pi_{\theta}(a|s)}{\pi_{\theta_{old}}(a|s)}$ is the ratio of likelihood between the two policies, $A_t$ is the generalized advantage estimation (GAE) proposed in~\cite{schulman2015high}, and $\beta$ is a dynamically adjusted coefficient to constrain the KL-divergence. We use a discount factor $\gamma = 0.99$ across all of our training runs.

The FleX physics simulator that we use can simulate a large collection of particles while running on a single GPU. To take advantage of this compute power we run many rollouts at once on the GPU.  Specifically, we simulate a $7 \times 7$ grid of individual rollouts in FleX, all of which are controlled by the actions generated from the policy at the current iteration. In total, we sample $49,000$ simulation steps and use $64$ samples per batch and use $10$ epochs in every training iteration.

\subsection{Simulation}

\label{section:simulation}
We simulate a one-way coupling system where a kinematic tool is controlled to interact with amorphous material and the material does not affect the motion of the tool. With FleX, we use position-based dynamics (PBD) \cite{macklin2014unified} to simulate the material as a collection of $P$ particles with positions $\mathbf{x}_{mat}^{p}$. FleX provides a rich set of materials and is highly efficient, making it well suited for our task.

The manipulation tool is simulated as a kinematic object with a state vector $\mathbf{s}=\left[\mathbf{q},\dot{\mathbf{q}}\right]$ where $\mathbf{q}=\left[x,y,z,\phi,\psi,\theta \right]^T$ are the position and orientation in Euler angles of the tool and $\dot{\mathbf{q}}$ are the corresponding time derivatives. To control the tool, a target position and orientation vector $\hat{\mathbf{q}}$ is provided by the policy at each time step, and an acceleration vector $\ddot{\mathbf{q}}$ is computed using a PD-controller as $\ddot{\mathbf{q}} = k_p(\hat{\mathbf{q}}-\mathbf{q})-k_d\dot{\mathbf{q}}$, where $k_p$ and $k_d$ are the gain and damping terms. Using the acceleration vector, the state of the tool is explicitly integrated as 

\begin{equation}
\begin{bmatrix}\mathbf{q}_{t+1} \\ \dot{\mathbf{q}}_{t+1}\end{bmatrix} = \begin{bmatrix}\mathbf{q}_{t} \\ \dot{\mathbf{q}}_{t}\end{bmatrix} + \Delta{t}\begin{bmatrix}\dot{\mathbf{q}}_{t} \\ \ddot{\mathbf{q}}_{t}\end{bmatrix},
\end{equation}
where $\Delta{t}$ is the simulation time step. By controlling the tool through this higher-order integration process instead of directly modifying its state, we obtain smoother motion of the tool.

In our experiments, we use three types of materials for demonstrating our learning algorithm: granular material, visco-plastic material, and viscous fluid material. Simulating granular and viscous fluid materials can be done directly in FleX using existing simulation models \cite{macklin2014unified}. However, the existing materials in FleX does not support effective simulation of visco-elastic materials, thus we created our own visco-plastic model in FleX.
\begin{figure*}[t!]
    \centering
    \includegraphics[width=\high]{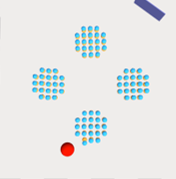}
    \hfill
    \includegraphics[width=\high]{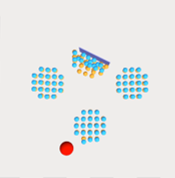}
    \hfill
    \includegraphics[width=\high]{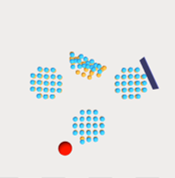}
    \hfill
    \includegraphics[width=\high]{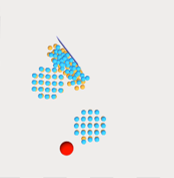}
    \hfill
    \includegraphics[width=\high]{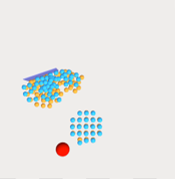}
    \hfill
    \includegraphics[width=\high]{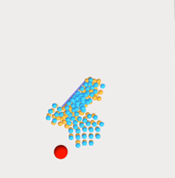}
    \hfill
    \includegraphics[width=\high]{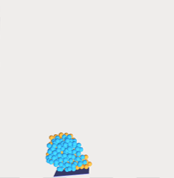}
    \caption{Visco-plastic gathering using a scraper. }
    \label{fig:plastic_gather}
\end{figure*}

\begin{figure*}[t!]
    \centering
    \includegraphics[width=\high]{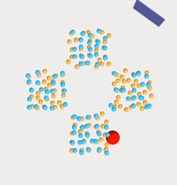}
    \hfill
    \includegraphics[width=\high]{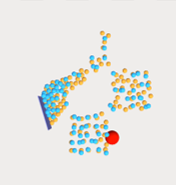}
    \hfill
    \includegraphics[width=\high]{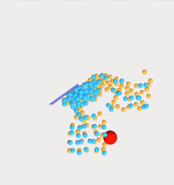}
    \hfill
    \includegraphics[width=\high]{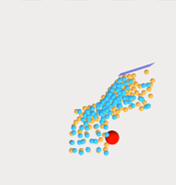}
    \hfill
    \includegraphics[width=\high]{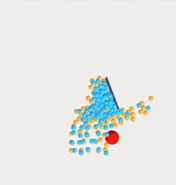}
    \hfill
    \includegraphics[width=\high]{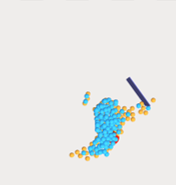}
    \hfill
    \includegraphics[width=\high]{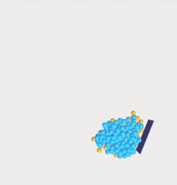}
    \caption{Granular gathering.}
    \label{fig:granular_gather}
\end{figure*}

\begin{figure*}[t!]
    \centering
    \includegraphics[height=\high]{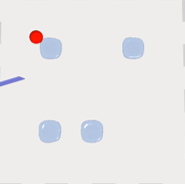}
    \hfill
    \includegraphics[height=\high]{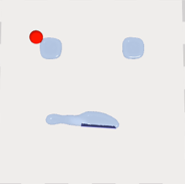}
    \hfill
    \includegraphics[height=\high]{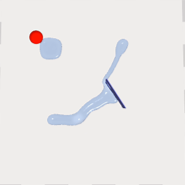}
    \hfill
    \includegraphics[height=\high]{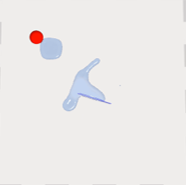}
    \hfill
    \includegraphics[height=\high]{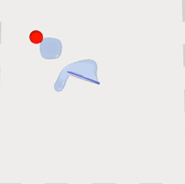}
    \hfill
    \includegraphics[height=\high]{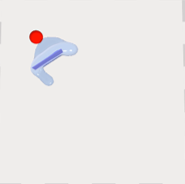}
    \hfill
    \includegraphics[height=\high]{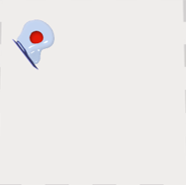}
    \caption{Viscous fluid gathering.}
    \label{fig:fluid_gather}
\end{figure*}
\subsubsection{Visco-plastic Material Modeling}

We want the visco-plastic material to be able to recover from small deformations, not returning to its former shape if it undergoes a large deformation. Also, when experiencing extreme stretching, we expect the material to split, and when compressed, separate pieces of material should merge. We simulate these behaviours by dynamically creating, adjusting, and removing a list of virtual springs $s_{i,j}$ with rest lengths $\mathcal{L}_{i,j}$ among the particles, where $i$ and $j$ denote the indices of the particles that the spring connects. We define the following values $d_{m}$, $d_{b}$, $r_{c}$, $r_{s}$, where $d_m$ and $d_b$ are distance thresholds that determine merging or breaking materials, $r_{c}$ and $r_s$ are ratio thresholds that determine when to plastically deform. Algorithm \ref{alg:plastic_modeling} describes how the rest lengths of the springs are updated and how the springs are created or removed. After determining the states of the springs, we simulate them in FleX as distance constraints.

\begin{algorithm}[tb]
   \caption{Visco-plastic Material Modeling}
   \label{alg:plastic_modeling}
\begin{algorithmic}[1]
    \FOR{Each spring $s_{i,j}$}
    \STATE $d =||\mathbf{x}_{mat}^i-\mathbf{x}_{mat}^j||$
    \IF{$d<d_b$ and not cut by tool}
    \STATE $\mathcal{L}_{i,j} = \text{RestLengthUpdate}(d,\mathcal{L}_{i,j})$
    \ELSE
    \STATE Delete $s_{i,j}$
    \ENDIF
    \ENDFOR
    
    \FOR{Every pair of particles with position $\mathbf{x}_{mat}^i$ and $\mathbf{x}_{mat}^j$}
    \STATE $d =||\mathbf{x}_{mat}^i-\mathbf{x}_{mat}^j||$
    \IF{$d<d_m$ and not cut by tool}
    \IF{$s_{i,j}$ does not exist}
    \STATE Create $s_{i,j}$ with $\mathcal{L}_{i,j} = d$
    \ENDIF
    \ENDIF
    \ENDFOR
\end{algorithmic}
\end{algorithm}

\begin{algorithm}[tb]
   \caption{RestLengthUpdate}
   \label{alg:rest_length_update}
\begin{algorithmic}[1]
    \STATE {\bfseries Input:} Current particle distance $d$, \\Current spring rest length $d_{rest}$
    \STATE $r_t = \frac{d}{d_{rest}}$
    \IF{$r_t \leq r_c$ or $r_t \geq r_s$}
    \STATE {\bfseries return: } $d$
    \ELSE
    \STATE {\bfseries return: } $d_{rest}$
    \ENDIF
\end{algorithmic}
\end{algorithm}

%% file: results.tex
\section{Results}
\label{section:result}


We demonstrate three different material manipulation tasks: gathering, spreading, and flipping. Table~\ref{table:policy_table} lists the different tasks and their training time. Note that even with GPU acceleration, training usually requires a full day or more. We discuss each of these tasks below, and animations are included in the accompanying video.

\begin{figure*}[h!]
    \centering
    \includegraphics[height=\high]{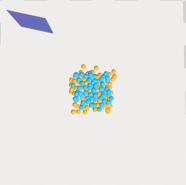}
    \hfill
    \includegraphics[height=\high]{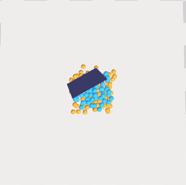}
    \hfill
    \includegraphics[height=\high]{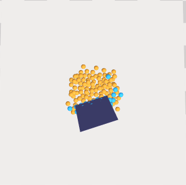}
    \hfill
    \includegraphics[height=\high]{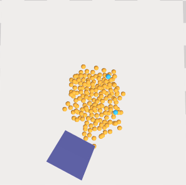}
    \hfill
    \includegraphics[height=\high]{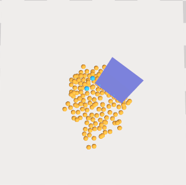}
    \hfill
    \includegraphics[height=\high]{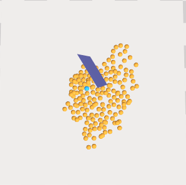}
    \hfill
    \includegraphics[height=\high]{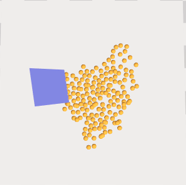}
    \caption{Granular material spreading. Particles that are yellow are touching the table's surface, and cyan particles are not touching.}
    \label{fig:spread_granular}
\end{figure*}

\begin{figure*}[h!]
    \centering
    \includegraphics[height=\high]{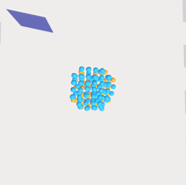}
    \hfill
    \includegraphics[height=\high]{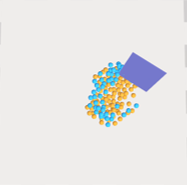}
    \hfill
    \includegraphics[height=\high]{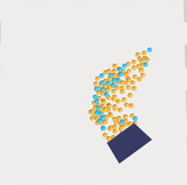}
    \hfill
    \includegraphics[height=\high]{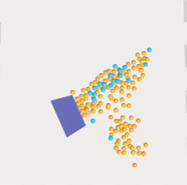}
    \hfill
    \includegraphics[height=\high]{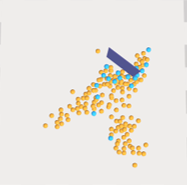}
    \hfill
    \includegraphics[height=\high]{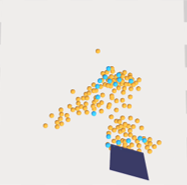}
    \hfill
    \includegraphics[height=\high]{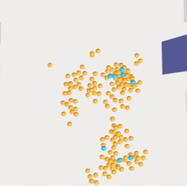}
    \caption{Visco-plastic material spreading.}
    \label{fig:spread_plastic}
\end{figure*}
\subsection{Gathering}

In this task, various clusters of material are placed across the surface of a flat table. The task is to slide a hibachi scraper across the table in order to gather the material at a specified goal position.  Figures~\ref{fig:plastic_gather}, \ref{fig:granular_gather}, and \ref{fig:fluid_gather} show sequences of applying a policy to gather visco-plastic, granular, and viscous fluid materials. 

The material observation space $O_{mat}$ consists of two images: a density map of the material, and a goal location map showing the location of the goal. For the motion of the scraper, we constrain its pose so that it can only slide on the table surface and rotate around the vertical axis. The translational states $\mathbf{x}_{tool}$ and $\dot{\mathbf{x}}_{tool}$ are $2D$ vectors storing the tool position and velocity on the plane of the table. The rotational states $\mathbf{r}_{tool}$ and $\dot{\mathbf{r}}_{tool}$ are the rotation angle and angular velocity about the axis orthogonal to the table plane. Similarly, the action is a $3D$ vector representing the $2D$ target position and $1D$ rotation angle in the tool's local coordinate frame.

We design the reward function for the gathering task into two stages: 1) encourage movement of the tool to a pose that will allow later motions to easily sweep material to the goal, 2) encourage moving the particles to the desired goal location. Given the goal position on the surface $\hat{\mathbf{x}}$, we compute a value that measures the distance from all particles to the goal as $d_{part}(\mathbf{x}_{mat}^i) = (c_1+||\mathbf{x}_{mat}^i-\hat{\mathbf{x}}||)^2$ and get the position of the particle that is farthest to the goal $\mathbf{x}_{far} = \argmax d_{part}(\mathbf{x}_{mat}^i)$. We then compute the distance between the tool position $\mathbf{x}_{tool}$ and the farthest particle as $d_{tool} = ||\mathbf{x}_{tool}-\mathbf{x}_{far}||$. We compute $r_{p} = c_2 + w_1(\Delta_t d_{part}(\mathbf{x}_{far}))+w_2\sum_{p=0}
^P||\Delta_t \mathbf{x}_{mat}^p|| + w_4\mathbbm{1}(d_{part}(\mathbf{x}_{far})<c_3)$, where $\Delta_t \cdot$ denotes the difference of the quantities at timestep $t$  and $t+1$. This reward term encourages moving particles to the goal. We also compute $r_{tool} = -w_3d_{tool}$ that encourages the tool to move towards the farthest particle. The final reward defined below is chosen from the two terms based on a dynamically determined threshold $d_{thr} =\max( \frac{c_4}{d_{part}(\mathbf{x}_{far})},c_5)$
\begin{equation}
r_t = \begin{cases} r_p & d_{tool} < d_{thr} \\
r_{tool} & \text{otherwise}\end{cases}.
\end{equation}
To encourage gathering various number of clusters of materials to the target location, we use a curriculum during training. We start training the policy with a single cluster of material placed on the table. If a rollout during the training produces a total reward greater than a threshold, we advance the training to a new curriculum by adding another cluster of material at the start of the rollouts. We keep adding clusters until the policy has been trained with four clusters.

\begin{table}[t]
\caption{Training data for each task}
\vskip -0.2in
\hskip -0.2in
\begin{center}
\begin{small}
\begin{sc}
\begin{tabular}{lccc}
\toprule
Tasks&Particles&Time Steps& Training Time\\ 
\midrule
Gathering&50-200&$\sim30M$&$\sim 30h$\\
Spreading&180&$\sim35M$&$\sim 39h$\\
Tossing&42&$\sim24M$&$\sim 22h$\\
\bottomrule
\end{tabular}
\end{sc}
\end{small}
\end{center}
\vskip -0.1in
\label{table:policy_table}
\end{table}

We trained two different gathering policies, one with granular material and another with visco-plastic material. The policy trained with granular material was unable to learn to gather more than a single pile. Training with visco-plastic material instead produced a policy that is able to sweep up four different material piles. Moreover, this same policy is also able to gather both granular and viscous fluid material. 
\begin{figure*}[h!]
    \centering
    \includegraphics[height=2cm]{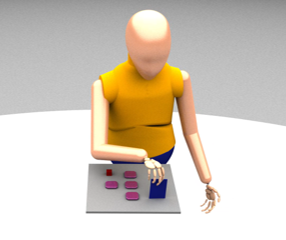}
    \hfill
    \includegraphics[height=2cm]{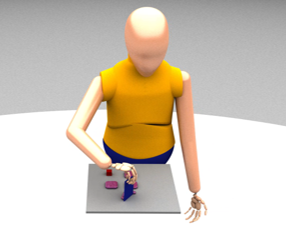}
    \hfill
    \includegraphics[height=2cm]{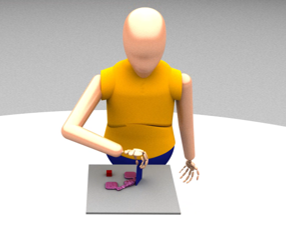}
    \hfill
    \includegraphics[height=2cm]{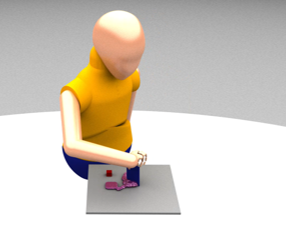}
    \hfill
    \includegraphics[height=2cm]{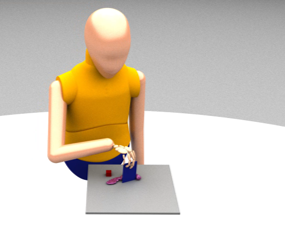}
    \hfill
    \includegraphics[height=2cm]{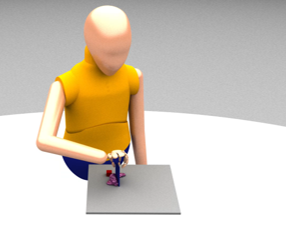}
    \hfill
    \includegraphics[height=2cm]{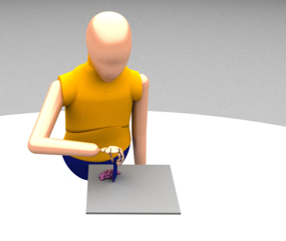}
    \caption{Character visco-plastic gathering.}
    \label{fig:character_gather}
\end{figure*}

\begin{figure*}[h!]
    \centering
    \includegraphics[height=2.4cm]{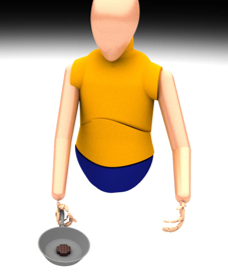}
    \hfill
    \includegraphics[height=2.4cm]{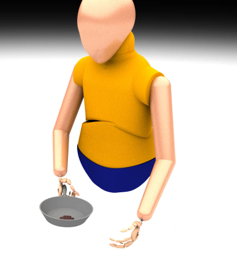}
    \hfill
    \includegraphics[height=2.4cm]{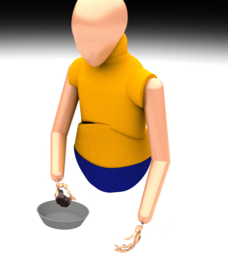}
    \hfill
    \includegraphics[height=2.4cm]{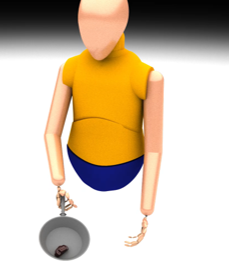}
    \hfill
    \includegraphics[height=2.4cm]{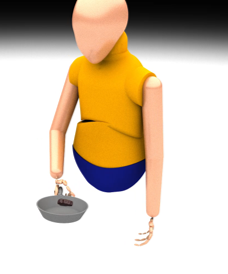}
    \hfill
    \includegraphics[height=2.4cm]{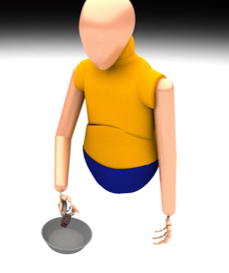}
    \hfill
    \includegraphics[height=2.4cm]{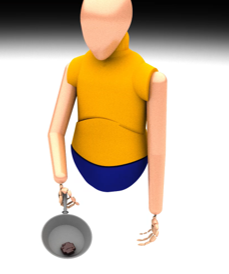}
    \caption{Character flipping.}
    \label{fig:character_flipping}
\end{figure*}

\subsection{Spreading}

In the spreading task, we consider the opposite problem to the gathering task: we take an initially tight pile of material and we spread it evenly over the table's surface. We show that a policy could spread visco-plastic materials in Figure ~\ref{fig:spread_plastic} and spread granular materials in Figure ~\ref{fig:spread_granular}.

In the material observation space $O_{mat}$, we provide both a density map and a height map for the material relative to the tool's coordinate frame. For the tool state, $\mathbf{x}_{tool}$ and $\dot{\mathbf{x}}_{tool}$ are each a $3D$ vector representing the position and velocity in all three dimensions. $\mathbf{r}_{tool}$ and $\dot{\mathbf{r}}_{tool}$ are $2D$ vectors that represents the Euler angles and angular velocities about the vertical axis and the axis aligned with the bottom edge of the scraper. The action space contains the $3D$ position of the tool as well as its rotation around the axis orthogonal to the table surface. To enforce the scrapper to push against the material during spreading, we tilt the scraper forward in the direction of its motion. 


The reward function for the spreading task is designed as a combination of several terms: $r_t =w_1r_{m}+w_2r_{hc}+w_3r_{h}+w_4r_{o}$. The height count reward $r_{hc}$ measures the change in the number of occupied cells of the height maps of particles between two consecutive states:
\begin{equation}
r_{hc} = sum(\mathbbm{1}(H_{t+1})>0))-sum(\mathbbm{1}(H_{t})>0)),
\end{equation}
where $\mathbbm{1}(H(\mathbf{x})>0))$ counts the number of grid cells that has non-zero particle occupancy. The mean height reward $r_{h}$ measures the change in the average height of the material between two states: 
\begin{equation}
r_{h} = \bar{y}_{t}-\bar{y}_{t+1},
\end{equation}
where $\bar{y}$ is the average height of all the particles. To encourage movement of particles above a certain height and penalized movement of particles on the table surface, we compute the particle movement reward $r_m$ as:
\begin{equation}
r_{m} = \frac{1}{n}\sum_{i=1}^n \begin{cases}||\mathbf{x}_{t+1}^i-\mathbf{x}_t^i|| & y_{t+1}^i>c_{min} \\
-k||\mathbf{x}_{t+1}^i-\mathbf{x}_t^i|| & otherwise\end{cases}
\end{equation}
where $c_{min}$ is a constant threshold.


To avoid particles being pushed outside the edge of the table, we use a regularization term $r_{o}$ penalizing the outlier particles as:
\begin{equation}
r_o = -\sum_{i=1}^P max(||\mathbf{x}_{mat}^i||-c_{rad},0),
\end{equation}
where $c_{rad}$ is the radius of the specified interior region.

Similar to the gathering task, we trained two different spreading policies, one using granular material and the other using visco-plastic material. Both policies can accomplish the spreading task successfully.

\subsection{Flipping}

In the third task, we demonstrate controlling a pan to toss and flip a disk of material in the air. See Figure~\ref{fig:character_flipping} for a character that uses such a policy. The observation of the material for this task is a single image of the material height map relative to the tool. For the state of the tool, we observe the $3D$ position and velocity as well as the Euler angle and angular velocity around the horizontal axis. In addition to these, we have a $3D$ vector representing the angular velocity of the material that is observed by the policy. The action for the vector is a $4D$ vector with $3$ translation and $1$ rotation degree-of-freedom to control the position and tilting of the frying pan.

The reward function defined for this task is a combination of two terms as $r = w_hr_{h}+w_{av}r_{av}$. By computing the minimum vertical distance of particles to the pan $d_{ymin} = \min_{i \in \{0, \dots, P\}}(y^i-y^{tool})$, the first term $r_h$ is defined as 
\begin{equation}
r_h = \begin{cases}c_1+w_{1} d_{ymin} & d_{ymin}>0\\
w_2d_{ymin} & otherwise
\end{cases}
\end{equation}
In addition to encouraging height, we also want to encourage the rotating motion of the material so that it can be flipped in the air. We compute the angular velocity vector of the material $\boldsymbol{\omega} = \frac{1}{n}\sum_{i=0}^n{\frac{(\mathbf{x}_{mat}^i-\bar{\mathbf{x}}) \times \mathbf{v}_i}{||\mathbf{x}_{mat}^i-\bar{\mathbf{x}}||^2}}$, where $\bar{\mathbf{x}}$ is the center-of-mass of all the particles, and $\mathbf{v}_i$ is the velocity of the $i$th particle. We then define the angular velocity reward term as
\begin{equation}
r_{av} =  \begin{cases}w_3clip(\boldsymbol{\omega} \cdot [1,0,0]^T,c_{\omega min},c_{\omega max}) & d_{ymin}>c_2 \\ 0 & otherwise \end{cases}.
\end{equation}

Because we wanted the material being flipped to hold together, we used an elastic material for the disk. We can tune the reward function for this task to encourage flipping in either of two directions. We find that roughly half of the training runs for this task create successful policies. The initial weights of the neural net appear to have an effect on the success of the policy training. We do not find this to be surprising, since there is substantial variation in the policy training outcomes that are reported in the reinforcement learning literature. All the constants and weights of all the experiments are included in Table \ref{table:weights}.
\begin{table}[h]
\caption{Parameter Choices}
\vskip -0.2in
\hskip -0.2in
\begin{center}
\begin{small}
\begin{sc}
\begin{tabular}{lccccc}

\midrule
Gathering&$c_1=10$&$c_2=0.03$&$c_3=1$&$c_4=4$&\\
&$c_5=1$&$w_1=1$&$w_2=5$&$w_3=0.01$&\\
&$w_4=1$&$d_{thr}=1.3$&&&\\
Spreading&$w_1=0.1$&$w_2=0.1$&$w_3=50$&$w_4=0.001$&\\
&$c_{min}=0.2$&$c_{rad}=3.2$&&&\\
Flipping&$w_h=0.1$&$w_{av}=1$&$w_1=10$&$w_2=0.1$&\\
&$w_3=4$&$c1=0.1$&$c_2=0.5$&$c_{\omega min}=-1$&\\
&$c_{\omega max}=1$&&&&\\
\bottomrule
\end{tabular}
\end{sc}
\end{small}
\end{center}

\label{table:weights}
\end{table}
\begin{figure*}[h!]
    \centering
    \includegraphics[width=0.35\textwidth]{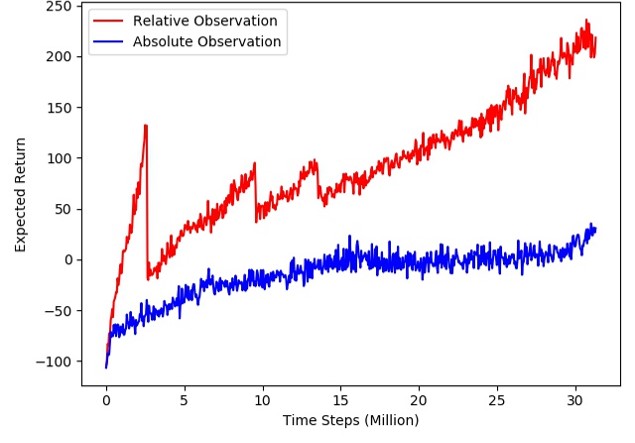}
    \hspace{15mm}
    \includegraphics[width=0.35\textwidth]{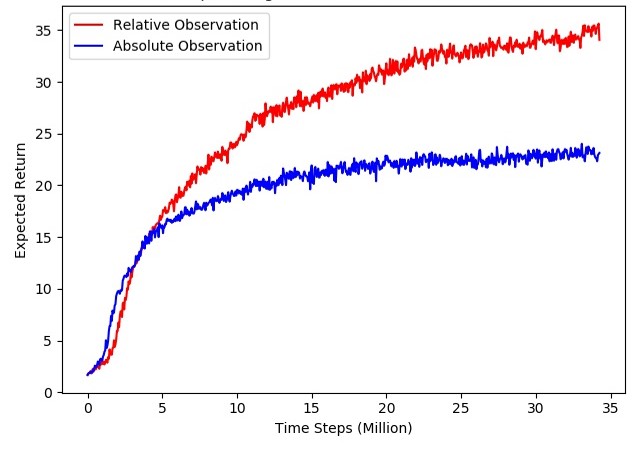}
    \vspace{-2mm}
    \caption{
    Comparison of relative and absolute observation for plastic gathering (left) and spreading (right) tasks. The three sharp drops of the red curve in the gathering task are when the curriculum has added another block of material. 
    }
    \label{fig:comp_learn}
\end{figure*}

\subsection{Character Animation}

To generate an animation of a character controlling the tool that manipulates the amorphous material, we use constrained optimization to solve a inverse kinematics (IK) problem. The desired solution of the IK is for the character to hold the tool and move it to track the pose of the tool that is generated by the manipulation policy. We solve:
\begin{equation}
    \min\limits_{\mathbf{q}_{char}'} ||\mathbf{q}_{char}'-\mathbf{q}_{char}||
\end{equation}

subject to 
\begin{equation}
\begin{aligned}
\mathbf{x}_{tool}(\mathbf{q}_{char}') &= \hat{\mathbf{x}}_{tool}\\
\mathbf{r}_{tool}(\mathbf{q}_{char}') &= \hat{\mathbf{r}}_{tool}\\
C_{joint}(\mathbf{q}_{char}') &\geq 0
\end{aligned},
\end{equation}
where $\mathbf{q}_{char}'$ is the solution pose of the character in generalized coordinate, and $\mathbf{q}_{char}$ is the current character generalized coordinate, $\mathbf{x}_{tool}(\mathbf{q}_{char}')$ and $\mathbf{r}_{tool}(\mathbf{q}_{char}')$ are the position and orientation of the tool held by the character, and $\hat{\mathbf{x}}_{tool}$ and $\hat{\mathbf{r}}_{tool}$ are the target position and orientation of the tool generated from the manipulation policy. Lastly, $C_{joint}(\mathbf{q}_{char}')$ are the realistic joint limit constraints proposed in \cite{jiang2018data}. Results of a character controlling a tool to manipulate materials are shown in Figures \ref{fig:teaser}, \ref{fig:character_gather} and \ref{fig:character_flipping}.

\begin{figure*}[h!]
    \centering
    \subfigure[]{
        \includegraphics[width=1.5in]{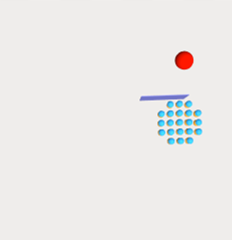}
        \hfill
        \includegraphics[width=1.5in]{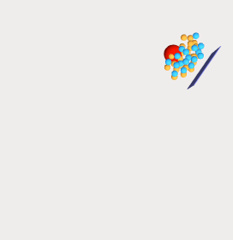}
        \label{subfig:gather_comp}
    }
    \subfigure[]{
        \includegraphics[width=1.5in]{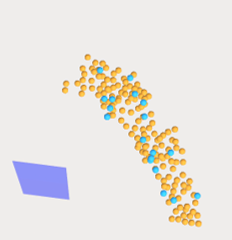}
        \hfill
        \includegraphics[width=1.5in]{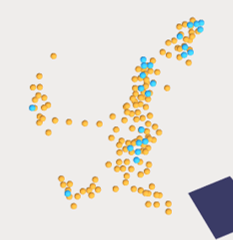}
        \label{subfig:spreading_comp}
    }
    \caption{Comparison of plastic material gathering and spreading. (a) shows the final state of a gathering policy using absolute observation and relative observation. (b) shows the similar states for the spreading policy.
    }
    \label{fig:comp_gathering_spreading}
\end{figure*}

%% file: ablation_study.tex
\subsection{Comparing Observation Coordinate Frames}

In this section we examine the importance of the coordinate frame of the image observations.  In particular, we compare an absolute coordinate frame to a coordinate frame that is relative to the tool's position and orientation. We compare the training results from both a spreading task and a gathering task on the visco-plastic material as described above. For the gathering task, the policy trained with relative observation completes the first curriculum within $3M$ time steps and eventually be able to gather up to $4$ piles of particles in $30M$ steps. For the policy trained with absolute observation, with $30M$ steps, the policy still failed to complete the gathering for the first curriculum, and the policy choose to control the scraper to reach to a certain pose and stays still for the rest of the rollout as shown in Figure \ref{subfig:gather_comp}. Their respective learning curve shown in Figure \ref{fig:comp_learn} shows the performance difference over training.

For the spreading policy, relative observations also perform better than the absolute observations. This can be seen in Figure \ref{subfig:spreading_comp}. The policy trained with relative observation is clearly able to spread the material in various directions, whereas the absolute observation trained policy is only able to spread the material along a single direction. With $\sim 35M$ time steps, we can see from Figure \ref{fig:comp_learn} that the training using relative observation outperforms the absolute observation throughout the training.


\subsection{Reward Ablation Study}
\label{section:ablation}
Designing a good reward function is crucial to the success of our algorithm. In this section, we investigate the importance of different reward terms used in each of the tasks (gather, spread, flip) by training a policy with the reward term removed and comparing the performance of the ablated reward policy to the baseline policy with all reward terms.
\subsubsection{Gathering}
For the gathering task, we study the importance of the following four reward terms: 
\begin{enumerate}
    \item $r_{tool}$ that encourages moving the tool to the farthest particle.
    \item $w_2\sum_{p=0}
^P||\Delta_t \mathbf{x}_{mat}^p||$ that encourages the movement of all particles.
    \item $w_4\mathbbm{1}(d_{part}(\mathbf{x}_{far})<c_3)$ that encourages particles to stay close to the vicinity of the goal.
    \item $w_1(\Delta_t d_{part}(\mathbf{x}_{far}))$ that encourages moving the farthest particle towards the goal.
\end{enumerate}
To compare the performance of each policy trained with different reward functions, we measure the farthest particle-to-goal distance across all particles at the last frame of the roll-out. We deploy each trained policy on $49$ rollouts with different random initializations and report the average performance of each policy. A comparison of the ablated policies to the main result (all reward terms) is shown in Table \ref{table:gathering_ablation}. This study shows that the reward term (4) is most important to the success of the learning, and a policy will not be able to complete the task without this term. Removing the other reward terms (1-3) also results in negative impacts on the policy performances, demonstrating the usefulness of these terms.
\begin{table}[h!]
\caption{Comparison of gathering policies with ablated reward terms. Numerical evaluation is the farthest particle-to-goal distance. Smaller value means better performance of the policy.}
\vskip -0.2in
\hskip -0.2in
\begin{center}
\begin{small}
\begin{sc}
\begin{tabular}{lcc}
\midrule
Full reward &$\textbf{1.06}m$\\
(1) $r_{tool}$ &$1.69m$\\
(2) $w_2\sum_{p=0}
^P||\Delta_t \mathbf{x}_{mat}^p||$ & $1.25\MakeLowercase{m}$\\
(3) $w_4\mathbbm{1}(d_{part}(\mathbf{x}_{far})<c_3)$ & $1.56\MakeLowercase{m}$\\
(4) $w_1(\Delta_t d_{part}(\mathbf{x}_{far}))$ &$6.63\MakeLowercase{m}$\\ 
\bottomrule
\end{tabular}
\end{sc}
\end{small}
\end{center}
\vskip -0.1in
\label{table:gathering_ablation}
\end{table}
\subsubsection{Spreading}
For the spreading task, we study how the absence of the following reward terms will affect the performance of a plastic material spreading policy.
\begin{enumerate}
    \item $r_m$ that encourages the movement of particles.
    \item $r_{hc}$ that encourages an increase in the number of occupied cells in the observed height map.
    \item $r_o$ that penalizes outlier particles that are pushed outside of the simulation region.
    \item $r_h$ that encourages a decreasing average height of the material 
\end{enumerate}
We use the number of occupied cells of the height map that covers the table at the end of the rollout to evaluate the performance of a policy. The importance of each reward term is shown in Table \ref{table:spreading_ablation}. The results show that both reward term (2) and (4) are critical to the success of the policy training. Interestingly, reward term (4), which rewards the task completion indirectly by rewarding lower material height, turns out to be more important than reward term (2), which is directly related to the task being trained for. This is possibly because the reward term (2) is more sparse as it counts the number of occupied cells, which may not change between steps, while reward (4) is a more continuous measure that provides a smoother shaped reward signal. On the other hand, reward terms (1) and (3) are helpful for the training, but removing them does not prevent the policy from learning to solve the task.
\begin{table}[h!]
\caption{Comparison of spreading policies with ablated reward terms. Numerical evaluation is the mean number of occupied cells. Larger value means better performance.}
\vskip -0.2in
\hskip -0.2in
\begin{center}
\begin{small}
\begin{sc}
\begin{tabular}{lcccc}
\midrule
Full reward &\textbf{336.12}\\
(1) $r_m$ &327.49\\
(2) $r_{hc}$ & 173.88\\
(3) $r_o$ &317.71\\
(4) $r_h$ &94.08\\
\bottomrule
\end{tabular}
\end{sc}
\end{small}
\end{center}
\vskip -0.1in
\label{table:spreading_ablation}
\end{table}
\subsubsection{Flipping}
For the flipping task, we study the importance of both the height rewarding term $r_h$ and the angular velocity rewarding term $r_{av}$ using the same process. As a result, both terms are important to the success of the policy. If $r_h$ is removed, the material will not be tossed high enough to consistently complete the flipping task. If $r_{av}$ is removed, the material will not be flipped in the desired direction. 

%% file: generalization.tex
\section{Generalization}

\begin{figure}[h]
    \centering
    \includegraphics[width=1.3in]{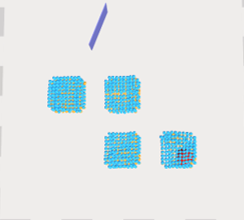}
    \hfill
    \includegraphics[width=1.3in]{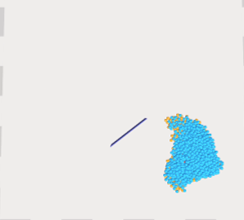}
    \caption{Initial and end frame of gathering policy on medium resolution visco-plastic material. This policy performs well.
    }
    \label{fig:gather_med_res}
\end{figure}

\begin{figure}[h]
    \centering
    \includegraphics[width=1.3in]{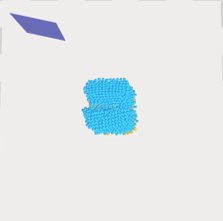}
    \hfill
    \includegraphics[width=1.3in]{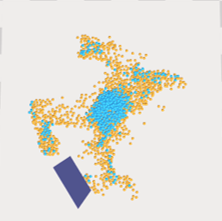}
    \caption{Initial and end frame of spreading policy on medium resolution visco-plastic material. This policy fails to spread the particles that start in the middle of the pile.
    }
    \label{fig:spread_med_res}
\end{figure}

\begin{figure}[h]
    \centering
    \includegraphics[width=1.3in]{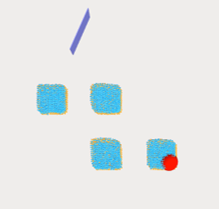}
    \hfill
    \includegraphics[width=1.3in]{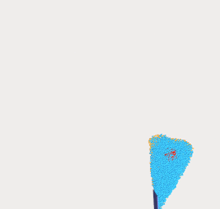}
    \caption{Initial and end frame of gathering policy on high resolution visco-plastic material. This policy performs well.
    }
    \label{fig:gather_high_res}
\end{figure}

\begin{figure}[h]
    \centering
    \includegraphics[width=1.3in]{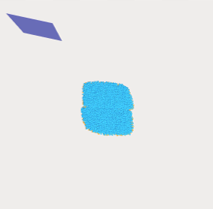}
    \hfill
    \includegraphics[width=1.3in]{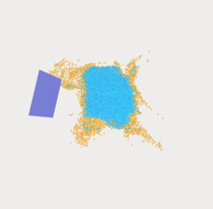}
    \caption{Initial and end frame of spreading policy on high resolution visco-plastic material. This policy fails to spread the particles that start in the middle of the pile.
    }
    \label{fig:spread_high_res}
\end{figure}

We further study how a trained policy will generalize to materials simulated with higher resolution. We mainly focus on the generalization of policies on the gathering and spreading tasks for the visco-plastic material. For each of the policy trained on low resolution material ($200$ particles for the gathering task and $180$ particles for the spreading task), we test it on a medium resolution and a high resolution material, where particle radius ratios are half and one-quarter compared to the training material particle size, respectively. For the gathering task, we used $1600$ particles in the medium resolution material and $5400$ particles in the high resolution material. For the spreading task, $1440$ particles and $4860$ particles are used in medium resolution material and high resolution material, respectively. Table \ref{table:resolution_comp} gives a comparison of using the trained policy on materials simulated with different resolution. The comparison uses the same metric described in \ref{section:ablation}. Figures \ref{fig:gather_med_res}, \ref{fig:spread_med_res},\ref{fig:gather_high_res},\ref{fig:spread_high_res} show the initial and end frame of a rollout for both gathering and spreading tasks with higher resolution materials. For the gathering task, our policy is able to achieve good performance for the high resolution materials. We note that the drop in the measured  performance for this task is partly due to the increased number of particles. For the spreading task, our policy can successfully generalize to the medium resolution material, while it failed in the high resolution material.
\begin{table}[t!]
\caption{Comparison of policy on different simulation resolution}
\vskip -0.2in
\hskip -0.2in
\begin{center}
\begin{small}
\begin{sc}
\begin{tabular}{lcccc}
\toprule
&Low Res&Medium Res&High Res\\ 
\midrule
Gathering&\textbf{1.06}&2.49&2.77\\
Spreading&336.12&\textbf{391.90}&96.04\\
\bottomrule
\end{tabular}
\end{sc}
\end{small}
\end{center}
\vskip -0.1in
\label{table:resolution_comp}
\end{table}



%% file: discussion.tex
\section{Discussion and Limitations}

For the gathering task, policies trained using the plastic material performed significantly better than those trained on granular material. For this task, we speculate that chasing down loose particles (more common with the granular material) may adversely impact the training. Differences in training success when using different materials suggests the intriguing idea of using a curriculum of different material properties when training for some tasks.

\begin{figure}[h]
    \centering
    \includegraphics[width=1.7in]{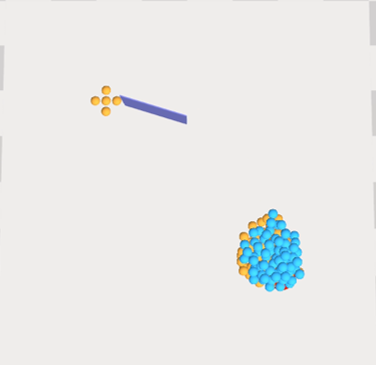}
    \caption{
    The scraper tries to reach the five particles but appears to hesitate before pushing them to the goal. 
    }
    \label{fig:outlier_gather}
\end{figure}

Although we were able to train successful policies for the gathering, spreading and flipping tasks, there is still room for improvements. One common type of failure case we observe for the gathering task is that the scraper sometimes appears to hesitate before resuming pushing of the material. An example rollout of this is shown in Figure \ref{fig:outlier_gather} where the scraper tries to reach the few particles on the upper left corner and fails to sweep them to the goal. This artifact might appear because the accurate position of the particles are not directly observed by the policy, introducing uncertainty to the policies behavior. For the spreading task, our policy does not achieve even spreading of the material to fully cover the entire table. Also, the materials are not guaranteed to stay connected afterwards. This could potentially be improved by using additional reward terms that encourage full coverage or connectedness of material during training. These properties would be important for reproducing many real-world tasks such as spreading paints.

Some of the limitations of our work are due to choices that we made in favor of faster policy training. The particle-based simulation in FleX is very efficient, yet is not able to model a large variety of materials. This leads us to the classic trade-off between better physical fidelity and higher computational cost. Simulators based on the Finite Element Methods (FEM) or the Material Point Methods (MPM) would provide more realistic simulations of visco-plastic materials or viscous fluid simulations.  Unfortunately, these higher quality simulators come at a higher computational cost.

Despite the choice of an efficient simulator, simulating amorphous material is still computationally expensive. As such, we use relatively low simulation resolution of the materials during training. In our work, we demonstrate preliminary but encouraging results in generalization to higher resolution materials. Applying domain adaptation techniques could further improve the generalization and allow efficient policy training for more complex problems.

Another limitation of our current work is that we do not take into account the range of motion of the character when training the motion of a given tool.  This means, for example, that a policy may be trained that rotates a hibachi scraper through 360 degrees or more while it gathers rice into a pile, which would require the character to re-grip the scraper.  We could prevent this by taking into account the character’s range of arm and hand motion during the policy training.  This would probably result in significantly higher training times, however.

%% file: conclusion.tex
\section{Conclusion}

We have developed a learning-based system for creating characters that can manipulate amorphous materials using tools like a frying pan or a scrapper. We demonstrate that by designing suitable observation spaces, action spaces, and reward functions of the learning problem, we can obtain successful and generalizable control policies that achieve interesting manipulation tasks despite the high computation cost of simulating non-rigid objects. One of our key techniques is to have the agent observe the environment relative to the current pose of the tool, which leads to better learning performance and generalization capabilities of the policy. We test our algorithm on three manipulation tasks: gathering materials to a certain position on the table, spreading the material evenly, and pan-flipping. We collectively show our system on multiple types of materials such as uncooked rice (granular), meat patty (plastic), and egg whites (viscous fluid).

In this work, we investigated manipulation tasks that involve relatively simple forms of tool usage: holding a pan or a scrapper, while humans can perform more sophisticated and intricate manipulation task with amorphous materials, such as making dumplings or peeling fruit. Many of these tasks involve dexterous manipulation, which notably increases the dimension and complexity of the action space. An interesting direction for future work is thus to expand our current learning system to handle these more challenging manipulation tasks. Another future direction is to include a more diverse set of materials with higher physical fidelity. To combat the potential additional cost in simulating these materials, one promising approach would be to use curriculum learning, where we train the policy with increasingly complicated materials throughout the learning. How to effectively choose the sequence of materials to form the curriculum in order to minimize the total training time would be an intriguing research problem.

%% file: main.bbl

\begin{thebibliography}{35}


\ifx \showCODEN    \undefined \def \showCODEN     #1{\unskip}     \fi
\ifx \showDOI      \undefined \def \showDOI       #1{#1}\fi
\ifx \showISBNx    \undefined \def \showISBNx     #1{\unskip}     \fi
\ifx \showISBNxiii \undefined \def \showISBNxiii  #1{\unskip}     \fi
\ifx \showISSN     \undefined \def \showISSN      #1{\unskip}     \fi
\ifx \showLCCN     \undefined \def \showLCCN      #1{\unskip}     \fi
\ifx \shownote     \undefined \def \shownote      #1{#1}          \fi
\ifx \showarticletitle \undefined \def \showarticletitle #1{#1}   \fi
\ifx \showURL      \undefined \def \showURL       {\relax}        \fi
\providecommand\bibfield[2]{#2}
\providecommand\bibinfo[2]{#2}
\providecommand\natexlab[1]{#1}
\providecommand\showeprint[2][]{arXiv:#2}

\bibitem[\protect\citeauthoryear{Akkaya, Andrychowicz, Chociej, Litwin, McGrew,
  Petron, Paino, Plappert, Powell, Ribas, et~al\mbox{.}}{Akkaya
  et~al\mbox{.}}{2019}]%
        {akkaya2019solving}
\bibfield{author}{\bibinfo{person}{Ilge Akkaya}, \bibinfo{person}{Marcin
  Andrychowicz}, \bibinfo{person}{Maciek Chociej}, \bibinfo{person}{Mateusz
  Litwin}, \bibinfo{person}{Bob McGrew}, \bibinfo{person}{Arthur Petron},
  \bibinfo{person}{Alex Paino}, \bibinfo{person}{Matthias Plappert},
  \bibinfo{person}{Glenn Powell}, \bibinfo{person}{Raphael Ribas},
  {et~al\mbox{.}}} \bibinfo{year}{2019}\natexlab{}.
\newblock \showarticletitle{Solving Rubik's Cube with a Robot Hand}.
\newblock \bibinfo{journal}{\emph{arXiv preprint arXiv:1910.07113}}
  (\bibinfo{year}{2019}).
\newblock


\bibitem[\protect\citeauthoryear{Andrews and Kry}{Andrews and Kry}{2012}]%
        {andrews2012policies}
\bibfield{author}{\bibinfo{person}{Sheldon Andrews} {and}
  \bibinfo{person}{Paul~G Kry}.} \bibinfo{year}{2012}\natexlab{}.
\newblock \showarticletitle{Policies for goal directed multi-finger
  manipulation}.
\newblock  (\bibinfo{year}{2012}).
\newblock


\bibitem[\protect\citeauthoryear{Bai and Liu}{Bai and Liu}{2014}]%
        {bai2014dexterous}
\bibfield{author}{\bibinfo{person}{Yunfei Bai} {and} \bibinfo{person}{C~Karen
  Liu}.} \bibinfo{year}{2014}\natexlab{}.
\newblock \showarticletitle{Dexterous manipulation using both palm and
  fingers}. In \bibinfo{booktitle}{\emph{2014 IEEE International Conference on
  Robotics and Automation (ICRA)}}. IEEE, \bibinfo{pages}{1560--1565}.
\newblock


\bibitem[\protect\citeauthoryear{Bai, Yu, and Liu}{Bai et~al\mbox{.}}{2016}]%
        {bai2016dexterous}
\bibfield{author}{\bibinfo{person}{Yunfei Bai}, \bibinfo{person}{Wenhao Yu},
  {and} \bibinfo{person}{C~Karen Liu}.} \bibinfo{year}{2016}\natexlab{}.
\newblock \showarticletitle{Dexterous manipulation of cloth}. In
  \bibinfo{booktitle}{\emph{Computer Graphics Forum}},
  Vol.~\bibinfo{volume}{35}. Wiley Online Library, \bibinfo{pages}{523--532}.
\newblock


\bibitem[\protect\citeauthoryear{Chebotar, Handa, Makoviychuk, Macklin, Issac,
  Ratliff, and Fox}{Chebotar et~al\mbox{.}}{2019}]%
        {chebotar2019closing}
\bibfield{author}{\bibinfo{person}{Yevgen Chebotar}, \bibinfo{person}{Ankur
  Handa}, \bibinfo{person}{Viktor Makoviychuk}, \bibinfo{person}{Miles
  Macklin}, \bibinfo{person}{Jan Issac}, \bibinfo{person}{Nathan Ratliff},
  {and} \bibinfo{person}{Dieter Fox}.} \bibinfo{year}{2019}\natexlab{}.
\newblock \showarticletitle{Closing the sim-to-real loop: Adapting simulation
  randomization with real world experience}. In \bibinfo{booktitle}{\emph{2019
  International Conference on Robotics and Automation (ICRA)}}. IEEE,
  \bibinfo{pages}{8973--8979}.
\newblock


\bibitem[\protect\citeauthoryear{Clegg, Yu, Tan, Liu, and Turk}{Clegg
  et~al\mbox{.}}{2018}]%
        {clegg2018learning}
\bibfield{author}{\bibinfo{person}{Alexander Clegg}, \bibinfo{person}{Wenhao
  Yu}, \bibinfo{person}{Jie Tan}, \bibinfo{person}{C~Karen Liu}, {and}
  \bibinfo{person}{Greg Turk}.} \bibinfo{year}{2018}\natexlab{}.
\newblock \showarticletitle{Learning to dress: Synthesizing human dressing
  motion via deep reinforcement learning}.
\newblock \bibinfo{journal}{\emph{ACM Transactions on Graphics (TOG)}}
  \bibinfo{volume}{37}, \bibinfo{number}{6} (\bibinfo{year}{2018}),
  \bibinfo{pages}{1--10}.
\newblock


\bibitem[\protect\citeauthoryear{Coumans and Bai}{Coumans and Bai}{2016}]%
        {coumans2016pybullet}
\bibfield{author}{\bibinfo{person}{Erwin Coumans} {and} \bibinfo{person}{Yunfei
  Bai}.} \bibinfo{year}{2016}\natexlab{}.
\newblock \showarticletitle{Pybullet, a python module for physics simulation
  for games, robotics and machine learning}.
\newblock \bibinfo{journal}{\emph{GitHub repository}} (\bibinfo{year}{2016}).
\newblock


\bibitem[\protect\citeauthoryear{Elliott and Cakmak}{Elliott and
  Cakmak}{2018}]%
        {elliott2018robotic}
\bibfield{author}{\bibinfo{person}{Sarah Elliott} {and} \bibinfo{person}{Maya
  Cakmak}.} \bibinfo{year}{2018}\natexlab{}.
\newblock \showarticletitle{Robotic cleaning through dirt rearrangement
  planning with learned transition models}. In \bibinfo{booktitle}{\emph{2018
  IEEE International Conference on Robotics and Automation (ICRA)}}. IEEE,
  \bibinfo{pages}{1623--1630}.
\newblock


\bibitem[\protect\citeauthoryear{Hu, Anderson, Li, Sun, Carr, Ragan-Kelley, and
  Durand}{Hu et~al\mbox{.}}{2019a}]%
        {hu2019difftaichi}
\bibfield{author}{\bibinfo{person}{Yuanming Hu}, \bibinfo{person}{Luke
  Anderson}, \bibinfo{person}{Tzu-Mao Li}, \bibinfo{person}{Qi Sun},
  \bibinfo{person}{Nathan Carr}, \bibinfo{person}{Jonathan Ragan-Kelley}, {and}
  \bibinfo{person}{Fr{\'e}do Durand}.} \bibinfo{year}{2019}\natexlab{a}.
\newblock \showarticletitle{DiffTaichi: Differentiable Programming for Physical
  Simulation}.
\newblock \bibinfo{journal}{\emph{arXiv preprint arXiv:1910.00935}}
  (\bibinfo{year}{2019}).
\newblock


\bibitem[\protect\citeauthoryear{Hu, Liu, Spielberg, Tenenbaum, Freeman, Wu,
  Rus, and Matusik}{Hu et~al\mbox{.}}{2019b}]%
        {hu2019chainqueen}
\bibfield{author}{\bibinfo{person}{Yuanming Hu}, \bibinfo{person}{Jiancheng
  Liu}, \bibinfo{person}{Andrew Spielberg}, \bibinfo{person}{Joshua~B
  Tenenbaum}, \bibinfo{person}{William~T Freeman}, \bibinfo{person}{Jiajun Wu},
  \bibinfo{person}{Daniela Rus}, {and} \bibinfo{person}{Wojciech Matusik}.}
  \bibinfo{year}{2019}\natexlab{b}.
\newblock \showarticletitle{ChainQueen: A real-time differentiable physical
  simulator for soft robotics}. In \bibinfo{booktitle}{\emph{2019 International
  Conference on Robotics and Automation (ICRA)}}. IEEE,
  \bibinfo{pages}{6265--6271}.
\newblock


\bibitem[\protect\citeauthoryear{Jiang and Liu}{Jiang and Liu}{2018}]%
        {jiang2018data}
\bibfield{author}{\bibinfo{person}{Yifeng Jiang} {and} \bibinfo{person}{C~Karen
  Liu}.} \bibinfo{year}{2018}\natexlab{}.
\newblock \showarticletitle{Data-driven approach to simulating realistic human
  joint constraints}. In \bibinfo{booktitle}{\emph{2018 IEEE International
  Conference on Robotics and Automation (ICRA)}}. IEEE,
  \bibinfo{pages}{1098--1103}.
\newblock


\bibitem[\protect\citeauthoryear{Kalashnikov, Irpan, Sampedro, Ibarz, Herzog,
  Jang, Quillen, Holly, Kalakrishnan, Vanhoucke, and Levine}{Kalashnikov
  et~al\mbox{.}}{2018}]%
        {DmitryQT}
\bibfield{author}{\bibinfo{person}{Dmitry Kalashnikov}, \bibinfo{person}{Alex
  Irpan}, \bibinfo{person}{Peter~Pastor Sampedro}, \bibinfo{person}{Julian
  Ibarz}, \bibinfo{person}{Alexander Herzog}, \bibinfo{person}{Eric Jang},
  \bibinfo{person}{Deirdre Quillen}, \bibinfo{person}{Ethan Holly},
  \bibinfo{person}{Mrinal Kalakrishnan}, \bibinfo{person}{Vincent Vanhoucke},
  {and} \bibinfo{person}{Sergey Levine}.} \bibinfo{year}{2018}\natexlab{}.
\newblock \showarticletitle{QT-Opt: Scalable Deep Reinforcement Learning for
  Vision-Based Robotic Manipulation}.
\newblock
\urldef\tempurl%
\url{https://arxiv.org/pdf/1806.10293}
\showURL{%
\tempurl}


\bibitem[\protect\citeauthoryear{Lee, Grey, Ha, Kunz, Jain, Ye, Srinivasa,
  Stilman, and Liu}{Lee et~al\mbox{.}}{2018}]%
        {lee2018dart}
\bibfield{author}{\bibinfo{person}{Jeongseok Lee}, \bibinfo{person}{Michael
  Grey}, \bibinfo{person}{Sehoon Ha}, \bibinfo{person}{Tobias Kunz},
  \bibinfo{person}{Sumit Jain}, \bibinfo{person}{Yuting Ye},
  \bibinfo{person}{Siddhartha Srinivasa}, \bibinfo{person}{Mike Stilman}, {and}
  \bibinfo{person}{Chuanjian Liu}.} \bibinfo{year}{2018}\natexlab{}.
\newblock \showarticletitle{Dart: Dynamic animation and robotics toolkit}.
\newblock \bibinfo{journal}{\emph{Journal of Open Source Software}}
  \bibinfo{volume}{3}, \bibinfo{number}{22} (\bibinfo{year}{2018}),
  \bibinfo{pages}{500}.
\newblock


\bibitem[\protect\citeauthoryear{Lee, Park, Lee, and Lee}{Lee
  et~al\mbox{.}}{2019}]%
        {lee2019scalable}
\bibfield{author}{\bibinfo{person}{Seunghwan Lee}, \bibinfo{person}{Moonseok
  Park}, \bibinfo{person}{Kyoungmin Lee}, {and} \bibinfo{person}{Jehee Lee}.}
  \bibinfo{year}{2019}\natexlab{}.
\newblock \showarticletitle{Scalable muscle-actuated human simulation and
  control}.
\newblock \bibinfo{journal}{\emph{ACM Transactions on Graphics (TOG)}}
  \bibinfo{volume}{38}, \bibinfo{number}{4} (\bibinfo{year}{2019}),
  \bibinfo{pages}{1--13}.
\newblock


\bibitem[\protect\citeauthoryear{Li, Wu, Tedrake, Tenenbaum, and Torralba}{Li
  et~al\mbox{.}}{2019}]%
        {li2018learning}
\bibfield{author}{\bibinfo{person}{Yunzhu Li}, \bibinfo{person}{Jiajun Wu},
  \bibinfo{person}{Russ Tedrake}, \bibinfo{person}{Joshua~B Tenenbaum}, {and}
  \bibinfo{person}{Antonio Torralba}.} \bibinfo{year}{2019}\natexlab{}.
\newblock \showarticletitle{Learning Particle Dynamics for Manipulating Rigid
  Bodies, Deformable Objects, and Fluids}. In \bibinfo{booktitle}{\emph{ICLR}}.
\newblock


\bibitem[\protect\citeauthoryear{Liu}{Liu}{2009}]%
        {liu2009dextrous}
\bibfield{author}{\bibinfo{person}{C~Karen Liu}.}
  \bibinfo{year}{2009}\natexlab{}.
\newblock \showarticletitle{Dextrous manipulation from a grasping pose}.
\newblock In \bibinfo{booktitle}{\emph{ACM SIGGRAPH 2009 papers}}.
  \bibinfo{pages}{1--6}.
\newblock


\bibitem[\protect\citeauthoryear{Liu and Hodgins}{Liu and Hodgins}{2018}]%
        {liu2018learning}
\bibfield{author}{\bibinfo{person}{Libin Liu} {and} \bibinfo{person}{Jessica
  Hodgins}.} \bibinfo{year}{2018}\natexlab{}.
\newblock \showarticletitle{Learning basketball dribbling skills using
  trajectory optimization and deep reinforcement learning}.
\newblock \bibinfo{journal}{\emph{ACM Transactions on Graphics (TOG)}}
  \bibinfo{volume}{37}, \bibinfo{number}{4} (\bibinfo{year}{2018}),
  \bibinfo{pages}{1--14}.
\newblock


\bibitem[\protect\citeauthoryear{Liu, Lehman, Molino, Such, Frank, Sergeev, and
  Yosinski}{Liu et~al\mbox{.}}{2018}]%
        {liu2018intriguing}
\bibfield{author}{\bibinfo{person}{Rosanne Liu}, \bibinfo{person}{Joel Lehman},
  \bibinfo{person}{Piero Molino}, \bibinfo{person}{Felipe~Petroski Such},
  \bibinfo{person}{Eric Frank}, \bibinfo{person}{Alex Sergeev}, {and}
  \bibinfo{person}{Jason Yosinski}.} \bibinfo{year}{2018}\natexlab{}.
\newblock \showarticletitle{An intriguing failing of convolutional neural
  networks and the coordconv solution}. In \bibinfo{booktitle}{\emph{Advances
  in Neural Information Processing Systems}}. \bibinfo{pages}{9605--9616}.
\newblock


\bibitem[\protect\citeauthoryear{Ma, Tian, Pan, Ren, and Manocha}{Ma
  et~al\mbox{.}}{2018}]%
        {ma2018fluid}
\bibfield{author}{\bibinfo{person}{Pingchuan Ma}, \bibinfo{person}{Yunsheng
  Tian}, \bibinfo{person}{Zherong Pan}, \bibinfo{person}{Bo Ren}, {and}
  \bibinfo{person}{Dinesh Manocha}.} \bibinfo{year}{2018}\natexlab{}.
\newblock \showarticletitle{Fluid directed rigid body control using deep
  reinforcement learning}.
\newblock \bibinfo{journal}{\emph{ACM Transactions on Graphics (TOG)}}
  \bibinfo{volume}{37}, \bibinfo{number}{4} (\bibinfo{year}{2018}),
  \bibinfo{pages}{1--11}.
\newblock


\bibitem[\protect\citeauthoryear{Macklin, M{\"u}ller, Chentanez, and
  Kim}{Macklin et~al\mbox{.}}{2014}]%
        {macklin2014unified}
\bibfield{author}{\bibinfo{person}{Miles Macklin}, \bibinfo{person}{Matthias
  M{\"u}ller}, \bibinfo{person}{Nuttapong Chentanez}, {and}
  \bibinfo{person}{Tae-Yong Kim}.} \bibinfo{year}{2014}\natexlab{}.
\newblock \showarticletitle{Unified particle physics for real-time
  applications}.
\newblock \bibinfo{journal}{\emph{ACM Transactions on Graphics (TOG)}}
  \bibinfo{volume}{33}, \bibinfo{number}{4} (\bibinfo{year}{2014}),
  \bibinfo{pages}{1--12}.
\newblock


\bibitem[\protect\citeauthoryear{Miller, Van Den~Berg, Fritz, Darrell,
  Goldberg, and Abbeel}{Miller et~al\mbox{.}}{2012}]%
        {miller2012geometric}
\bibfield{author}{\bibinfo{person}{Stephen Miller}, \bibinfo{person}{Jur Van
  Den~Berg}, \bibinfo{person}{Mario Fritz}, \bibinfo{person}{Trevor Darrell},
  \bibinfo{person}{Ken Goldberg}, {and} \bibinfo{person}{Pieter Abbeel}.}
  \bibinfo{year}{2012}\natexlab{}.
\newblock \showarticletitle{A geometric approach to robotic laundry folding}.
\newblock \bibinfo{journal}{\emph{The International Journal of Robotics
  Research}} \bibinfo{volume}{31}, \bibinfo{number}{2} (\bibinfo{year}{2012}),
  \bibinfo{pages}{249--267}.
\newblock


\bibitem[\protect\citeauthoryear{Mordatch, Popovi{\'c}, and Todorov}{Mordatch
  et~al\mbox{.}}{2012}]%
        {mordatch2012contact}
\bibfield{author}{\bibinfo{person}{Igor Mordatch}, \bibinfo{person}{Zoran
  Popovi{\'c}}, {and} \bibinfo{person}{Emanuel Todorov}.}
  \bibinfo{year}{2012}\natexlab{}.
\newblock \showarticletitle{Contact-invariant optimization for hand
  manipulation}. In \bibinfo{booktitle}{\emph{Proceedings of the ACM
  SIGGRAPH/Eurographics symposium on computer animation}}. Eurographics
  Association, \bibinfo{pages}{137--144}.
\newblock


\bibitem[\protect\citeauthoryear{Park, Hoshi, Mahajan, Kim, Erickson, Rogers,
  and Kemp}{Park et~al\mbox{.}}{2019}]%
        {park2019active}
\bibfield{author}{\bibinfo{person}{Daehyung Park}, \bibinfo{person}{Yuuna
  Hoshi}, \bibinfo{person}{Harshal~P Mahajan}, \bibinfo{person}{Ho~Keun Kim},
  \bibinfo{person}{Zackory Erickson}, \bibinfo{person}{Wendy~A Rogers}, {and}
  \bibinfo{person}{Charles~C Kemp}.} \bibinfo{year}{2019}\natexlab{}.
\newblock \showarticletitle{Active Robot-Assisted Feeding with a
  General-Purpose Mobile Manipulator: Design, Evaluation, and Lessons Learned}.
\newblock \bibinfo{journal}{\emph{arXiv preprint arXiv:1904.03568}}
  (\bibinfo{year}{2019}).
\newblock


\bibitem[\protect\citeauthoryear{Peng, Abbeel, Levine, and van~de Panne}{Peng
  et~al\mbox{.}}{2018}]%
        {peng2018deepmimic}
\bibfield{author}{\bibinfo{person}{Xue~Bin Peng}, \bibinfo{person}{Pieter
  Abbeel}, \bibinfo{person}{Sergey Levine}, {and} \bibinfo{person}{Michiel
  van~de Panne}.} \bibinfo{year}{2018}\natexlab{}.
\newblock \showarticletitle{DeepMimic: Example-Guided Deep Reinforcement
  Learning of Physics-Based Character Skills}.
\newblock \bibinfo{journal}{\emph{ACM Transactions on Graphics (Proc. SIGGRAPH
  2018)}} (\bibinfo{year}{2018}).
\newblock


\bibitem[\protect\citeauthoryear{Rajeswaran, Kumar, Gupta, Vezzani, Schulman,
  Todorov, and Levine}{Rajeswaran et~al\mbox{.}}{2017}]%
        {rajeswaran2017learning}
\bibfield{author}{\bibinfo{person}{Aravind Rajeswaran}, \bibinfo{person}{Vikash
  Kumar}, \bibinfo{person}{Abhishek Gupta}, \bibinfo{person}{Giulia Vezzani},
  \bibinfo{person}{John Schulman}, \bibinfo{person}{Emanuel Todorov}, {and}
  \bibinfo{person}{Sergey Levine}.} \bibinfo{year}{2017}\natexlab{}.
\newblock \showarticletitle{Learning complex dexterous manipulation with deep
  reinforcement learning and demonstrations}.
\newblock \bibinfo{journal}{\emph{arXiv preprint arXiv:1709.10087}}
  (\bibinfo{year}{2017}).
\newblock


\bibitem[\protect\citeauthoryear{Schenck and Fox}{Schenck and Fox}{2018}]%
        {schenck2018spnets}
\bibfield{author}{\bibinfo{person}{Connor Schenck} {and}
  \bibinfo{person}{Dieter Fox}.} \bibinfo{year}{2018}\natexlab{}.
\newblock \showarticletitle{Spnets: Differentiable fluid dynamics for deep
  neural networks}.
\newblock \bibinfo{journal}{\emph{arXiv preprint arXiv:1806.06094}}
  (\bibinfo{year}{2018}).
\newblock


\bibitem[\protect\citeauthoryear{Schenck, Tompson, Fox, and Levine}{Schenck
  et~al\mbox{.}}{2017}]%
        {schenck2017learning}
\bibfield{author}{\bibinfo{person}{Connor Schenck}, \bibinfo{person}{Jonathan
  Tompson}, \bibinfo{person}{Dieter Fox}, {and} \bibinfo{person}{Sergey
  Levine}.} \bibinfo{year}{2017}\natexlab{}.
\newblock \showarticletitle{Learning robotic manipulation of granular media}.
\newblock \bibinfo{journal}{\emph{arXiv preprint arXiv:1709.02833}}
  (\bibinfo{year}{2017}).
\newblock


\bibitem[\protect\citeauthoryear{Schulman, Moritz, Levine, Jordan, and
  Abbeel}{Schulman et~al\mbox{.}}{2015}]%
        {schulman2015high}
\bibfield{author}{\bibinfo{person}{John Schulman}, \bibinfo{person}{Philipp
  Moritz}, \bibinfo{person}{Sergey Levine}, \bibinfo{person}{Michael Jordan},
  {and} \bibinfo{person}{Pieter Abbeel}.} \bibinfo{year}{2015}\natexlab{}.
\newblock \showarticletitle{High-dimensional continuous control using
  generalized advantage estimation}.
\newblock \bibinfo{journal}{\emph{arXiv preprint arXiv:1506.02438}}
  (\bibinfo{year}{2015}).
\newblock


\bibitem[\protect\citeauthoryear{Schulman, Wolski, Dhariwal, Radford, and
  Klimov}{Schulman et~al\mbox{.}}{2017}]%
        {schulman2017proximal}
\bibfield{author}{\bibinfo{person}{John Schulman}, \bibinfo{person}{Filip
  Wolski}, \bibinfo{person}{Prafulla Dhariwal}, \bibinfo{person}{Alec Radford},
  {and} \bibinfo{person}{Oleg Klimov}.} \bibinfo{year}{2017}\natexlab{}.
\newblock \showarticletitle{Proximal policy optimization algorithms}.
\newblock \bibinfo{journal}{\emph{arXiv preprint arXiv:1707.06347}}
  (\bibinfo{year}{2017}).
\newblock


\bibitem[\protect\citeauthoryear{Todorov, Erez, and Tassa}{Todorov
  et~al\mbox{.}}{2012}]%
        {todorov2012mujoco}
\bibfield{author}{\bibinfo{person}{Emanuel Todorov}, \bibinfo{person}{Tom
  Erez}, {and} \bibinfo{person}{Yuval Tassa}.} \bibinfo{year}{2012}\natexlab{}.
\newblock \showarticletitle{Mujoco: A physics engine for model-based control}.
  In \bibinfo{booktitle}{\emph{2012 IEEE/RSJ International Conference on
  Intelligent Robots and Systems}}. IEEE, \bibinfo{pages}{5026--5033}.
\newblock


\bibitem[\protect\citeauthoryear{Wilson and Hermans}{Wilson and
  Hermans}{2019}]%
        {wilson2019learning}
\bibfield{author}{\bibinfo{person}{Matthew Wilson} {and}
  \bibinfo{person}{Tucker Hermans}.} \bibinfo{year}{2019}\natexlab{}.
\newblock \showarticletitle{Learning to Manipulate Object Collections Using
  Grounded State Representations}.
\newblock \bibinfo{journal}{\emph{arXiv preprint arXiv:1909.07876}}
  (\bibinfo{year}{2019}).
\newblock


\bibitem[\protect\citeauthoryear{Wu, Yan, Kurutach, Pinto, and Abbeel}{Wu
  et~al\mbox{.}}{2019}]%
        {wu2019learning}
\bibfield{author}{\bibinfo{person}{Yilin Wu}, \bibinfo{person}{Wilson Yan},
  \bibinfo{person}{Thanard Kurutach}, \bibinfo{person}{Lerrel Pinto}, {and}
  \bibinfo{person}{Pieter Abbeel}.} \bibinfo{year}{2019}\natexlab{}.
\newblock \showarticletitle{Learning to Manipulate Deformable Objects without
  Demonstrations}.
\newblock \bibinfo{journal}{\emph{arXiv preprint arXiv:1910.13439}}
  (\bibinfo{year}{2019}).
\newblock


\bibitem[\protect\citeauthoryear{Ye and Liu}{Ye and Liu}{2012}]%
        {ye2012synthesis}
\bibfield{author}{\bibinfo{person}{Yuting Ye} {and} \bibinfo{person}{C~Karen
  Liu}.} \bibinfo{year}{2012}\natexlab{}.
\newblock \showarticletitle{Synthesis of detailed hand manipulations using
  contact sampling}.
\newblock \bibinfo{journal}{\emph{ACM Transactions on Graphics (TOG)}}
  \bibinfo{volume}{31}, \bibinfo{number}{4} (\bibinfo{year}{2012}),
  \bibinfo{pages}{1--10}.
\newblock


\bibitem[\protect\citeauthoryear{Yu, Park, and Lee}{Yu et~al\mbox{.}}{2019}]%
        {yu2019figure}
\bibfield{author}{\bibinfo{person}{Ri Yu}, \bibinfo{person}{Hwangpil Park},
  {and} \bibinfo{person}{Jehee Lee}.} \bibinfo{year}{2019}\natexlab{}.
\newblock \showarticletitle{Figure Skating Simulation from Video}. In
  \bibinfo{booktitle}{\emph{Computer Graphics Forum}},
  Vol.~\bibinfo{volume}{38}. Wiley Online Library, \bibinfo{pages}{225--234}.
\newblock


\bibitem[\protect\citeauthoryear{Zhao, Zhang, Min, and Chai}{Zhao
  et~al\mbox{.}}{2013}]%
        {zhao2013robust}
\bibfield{author}{\bibinfo{person}{Wenping Zhao}, \bibinfo{person}{Jianjie
  Zhang}, \bibinfo{person}{Jianyuan Min}, {and} \bibinfo{person}{Jinxiang
  Chai}.} \bibinfo{year}{2013}\natexlab{}.
\newblock \showarticletitle{Robust realtime physics-based motion control for
  human grasping}.
\newblock \bibinfo{journal}{\emph{ACM Transactions on Graphics (TOG)}}
  \bibinfo{volume}{32}, \bibinfo{number}{6} (\bibinfo{year}{2013}),
  \bibinfo{pages}{1--12}.
\newblock


\end{thebibliography}
